\documentclass[preprint]{epl-noheader}

\title{Maximum-entropy theory of steady-state quantum transport}
\shorttitle{Maximum-entropy theory of quantum transport}
\author{P. Bokes\thanks{E-mail: \email{pb20@york.ac.uk}} and R. W. Godby}
\institute{ Department of Physics, University of York, Heslington, York
         YO10 5DD, United Kingdom }

\begin{document}

\maketitle

\begin{abstract}
We develop a new theoretical framework for describing
steady-state quantum transport phenomena, based on the general
maximum-entropy principle of non-equilibrium statistical mechanics.
The general form of the many-body density matrix is derived, which contains
the invariant part of the current operator that guarantees the non-equilibrium
and steady-state character of the ensemble. Several examples of the
theory are given, demonstrating the relationship of the present treatment
to the widely-used scattering-states occupation schemes at the level
of the self-consistent single-particle approximation.
\end{abstract}

During the last few years many {\it ab-initio} calculations have addressed  
the electronic structure of systems  with nonzero electrical 
current~\cite{Taylor,Lang,Hirose}. We will refer to these as the occupation 
scheme approaches~(OS) since in practise 
one occupies the right- and left-going scattering states up to two different 
electrochemical potentials, 
$\mu^R$ and $\mu^L$ respectively. The essential idea of the OS come
from Landauer's treatment of coherent transport in terms of
the transmission matrix of the conductor~\cite{Buttiker85}. Later,
Caroli {\it et al.}~\cite{Caroli}
and independently, Feuchtwang~\cite{Feuchtwang} developed a formal 
theory of tunnelling based on the technique of Keldysh non-equilibrium Green's 
functions~\cite{Keldysh} which could be extended to address coherent 
transport as well. Recently it has been shown~\cite{Taylor02} that these 
two approaches are indeed equivalent at the level of a single-particle 
approximation in the spirit of Kohn-Sham density-functional theory.
However, the latter approximation is very hard to justify. 
It is not clear what effective potential one should use; the use of the local
density approximation~(LDA) is a mere hope rather than a secure 
approximation. 
We also believe that a certain difficulty might lie in the formal search for 
steady-state non-equilibrium Green's functions using a unitary 
(Hamiltonian-driven) evolution for $t\rightarrow \infty$ from an undisturbed 
system. For example, when we adiabatically turn on an external field, 
the Keldysh technique predicts no change in temperature, in contradiction 
with statistical thermodynamics. We therefore believe that any 
alternative point of view is of great utility here. 

We build such an alternative theory using the generalised maximum-entropy
principle as established by Jaynes~\cite{Jaynes78}. Similar ideas
were heavily exploited in the development of the projector 
techniques for non-equilibrium statistical mechanics by 
Mori~\cite{Mori,Kubo59},  
yet  detailed application to concrete problems is not widespread. 
Of the few papers, let us mention those of Ng~\cite{Ng92}, 
and Heinonen and Johnson~\cite{Heinonen93}, who consider current-carrying 
ensembles, similar to ours. However, in these papers, the essential 
steady-state character is not considered. This results in
differences between the predictions of OS and maximum-entropy methods, in 
the linear-response regime, that are not present in our work. We give 
a coherent formalism that does not depend on the complexity of the system, 
i.e. without restriction to non-interacting particles or simple band-structure 
models. 

The statistical density matrix (DM), which represents an ensemble 
with known or controlled averages of given operators 
$\langle A_{i} \rangle = \mathrm{Tr}[\rho \hat{A}_{i} ]$, is obtained
by maximising the information entropy
$S[\hat{\rho}]= - \textrm{Tr} \left[ \hat{\rho} \log(\hat{\rho}) \right]$, 
subject to constraints on the traces of the above-mentioned 
operators~\cite{Jaynes78}. 
In the case of quantum transport, experiments suggest that 
for a given temperature, composition and total current 
we obtain a well-defined thermodynamic state (or, in the case of N-shaped 
{\it I-V} curves, a small number of states differing by applied bias voltage).

Firstly, the total energy is conserved. This constraint is associated with
the Lagrange multiplier $\beta$, corresponding to inverse temperature
for equilibrium - or near-equilibrium - systems.
Similarly, the number of electrons is conserved and on average is
given by the total positive charge in the background, {\it i.e.} atomic nuclei.
Therefore, a constraint on the number of particles is used, with the 
usual symbol $\mu$ for the related Lagrange multiplier.
The total current $I$ should be the next thermodynamical parameter of the 
theory. On the contrary, the vast majority of present approaches to quantum 
transport use the applied bias $\Delta V$ instead.  However, $\Delta V$ 
is not convenient for it is defined uniquely only between two ideal 
reservoirs, each being in equilibrium. These should never be a part 
of a practical calculation, to say nothing of the strongly non-local character 
of this quantity. In contrast, the current flowing through 
the system is represented by a simple operator and is well-defined 
even in the strongly non-equilibrium regime.
We use the symbol $A$ for the Lagrange multiplier accompanying  
the current constraint, and we later show that $A$ is universally related 
to $\Delta V$. 

Finally we impose the steady-state condition 
$\left[ \hat{\rho},\hat{H} \right] = 0$. 
For this to have a nontrivial solution, the system must be infinite 
along the direction of the current. Otherwise, the only steady state would
correspond to zero current. This is equally present in the Keldysh 
formalism, where one has to consider the limit of infinite size
first, and only afterwards can the time-evolution of the response to the 
turned-on transfer Hamiltonian go to infinity. To implement the steady-state 
constraint 
we write the steady-state condition in any complete set of states 
$\langle E,\alpha | \left[ \hat{\rho},\hat{H} 
\right] | E',\alpha' \rangle = 0 $ for all $E,E',\alpha,\alpha'$. 
This particular notation stresses the fact that 
we work with a continuum of eigenstates of $\hat{H}$, normalised to a delta 
function of energy~\footnote{The final results obtained do not depend on 
the particular choice of normalisation.}. 
The index $\alpha$ runs over the discrete set of degenerate states at 
energy $E$.  Each of these equations must be now 
guaranteed, with a separate Lagrange multiplier $\lambda_{\alpha,\alpha'}(E,E')$ 
and the expression in the functional to be maximised can be manipulated into
\begin{eqnarray}
        \int dE dE' \sum_{\alpha,\alpha'} \lambda_{\alpha',\alpha}(E',E)  
	\langle E,\alpha  | \left[ \hat{\rho},\hat{H} \right]
        | E',\alpha'  \rangle 
	 = \mathrm{Tr}\left[ \hat{\lambda}~[\hat{\rho},\hat{H}] \right]
        = \mathrm{Tr}\left[ \hat{\rho}~ \hat{L} \right],
        \label{time-constraint}
\end{eqnarray}
where we have introduced $\hat{L} = [\hat{H},\hat{\lambda}]$. This form is 
suitable for the variation with respect to the DM.

Collecting all the above terms we obtain the variational condition
\begin{eqnarray}
        \delta \left\{
        - \langle \log(\hat{\rho}) \rangle
        + (\Omega + 1) \langle \hat{1} \rangle
        - \beta \langle \hat{H} \rangle
        + \beta \mu \langle \hat{N} \rangle + \beta  
	A \langle \hat{I} \rangle \right. 
        \left.  - \beta \langle \hat{L} \rangle
        \right\} = 0. \quad
        \label{max-ent-principle}
\end{eqnarray}
The term $(\Omega + 1) \langle \hat{1} \rangle$
guarantees the normalisation of the DM. We also note that we have
deliberately introduced the parameter $\beta$ in the definition of
all the other multipliers so that the limit $\beta \rightarrow \infty$
can be conveniently studied. As a result of variation we obtain 
the stationary non-equilibrium DM
$ \hat{\rho} = \exp\{\Omega-\beta\hat{K}\}$, where 
$ \hat{K} = \hat{H} - \mu \hat{N} - A \hat{I} + \hat{L} $. 
The practicality of this expression relies on the knowledge of the
$\hat{L}$ operator. We obtain its form from the solution of
$\left[ \hat{\rho},\hat{H} \right]=0$, as an equation for $\hat{L}$. 
Expanding the DM in terms of $\hat{K}$, we see that this is equivalent to
$\left[ - A \hat{I} + \hat{L},\hat{H} \right] = 0$.
If we cast the last expression in the representation of the 
eigenstates of $\hat{H}$, it is seen that the role of $\hat{L}$ is to remove the
off-diagonal elements of the current operator. We shall show below that 
$\hat{L}$ should be of the form 
\begin{equation}
	L_{\alpha,\alpha'}(E,E') = I_{\alpha,\alpha'}(E,E')\left( A - \tilde{A} 
		\delta(E-E') \right),
	\label{L-straight-E}
\end{equation}
where $\tilde{A}$ is a finite constant, related to $A$ as 
$\tilde{A} = \pi \epsilon A$, with $\epsilon \sim 1/l$ an infinitely small 
energy inversely proportional to the length of the system. The result is 
a finite number, since by inspection of Eq.~(3) we deduce that $A \sim l$ 
due to $I \sim 1$.

To prove~(\ref{L-straight-E}) we set 
$\hat{Y}=-A\left[\hat{I},\hat{H}\right]$ and  write in 
the basis of $\{ |E,\alpha \rangle \}$ the equation for $\hat{L}$ as:
\begin{eqnarray}
        iY_{\alpha,\alpha'}(E,E') + iL_{\alpha,\alpha'}(E,E')(E'-E) &=& 0, 
\end{eqnarray}
for $E\neq E'$ we have
\begin{equation}
        L_{\alpha,\alpha'}(E,E') = \frac{iY_{\alpha,\alpha'}(E,E')}{i(E-E')}
	= AI_{\alpha,\alpha'}(E,E').   
        \label{offdiagL}
\end{equation}
Because $Y(E,E')$ is a result of a commutator, it is also proportional 
to $E-E'$. However, we need $L_{\alpha,\alpha'}(E,E')$
to be zero for $E=E'$ to keep the current at its given value and to satisfy
the fact that $\hat{L}$ is also result of a commutator with $\hat{H}$.
This is uniquely achieved by
\begin{equation}
        L_{\alpha,\alpha'}(E,E') = I_{\alpha,\alpha'}(E,E')\left( 
	A  - A \lim_{\epsilon \rightarrow 0^{+}} 
        \epsilon \pi \delta_{\epsilon}(E-E')\right), \label{L-derived-E}
\end{equation}
where
\begin{equation}
        \delta_{\epsilon}(E-E') = \frac{1}{\pi} 
        \frac{\epsilon}{(E-E')^2 + \epsilon^2},
\end{equation}
which manifestly satisfies both conditions, since for $E=E'$, 
$\pi \epsilon \delta_{\epsilon} = 1$.
This is the stated result~(\ref{L-straight-E}).
Eq.~(\ref{L-derived-E}) can be now written in basis-independent form as
\begin{equation}
	\hat{L} = A \left( \hat{I} - \hat{I}^{0} \right), \quad
	\hat{I}^{0} = \lim_{T\rightarrow \infty} \frac{1}{2T} 
	\int_{-T}^{T} \hat{I}(t) dt,
	\label{lambda-solution}
\end{equation}
where the operator $\hat{I}^{0}$ has the form of {\it the invariant part of 
the current operator} with respect to the time evolution, introduced 
by R. Kubo in the linear response theory~\cite{Kubo59},
where the time-dependence of the operator $\hat{I}(t)$ is determined by 
the Hamiltonian
$\hat{H}$. If we insert for the $\hat{L}$ in the stationary DM obtained from
Eq.~(\ref{max-ent-principle}) the solution Eq.~(\ref{lambda-solution}),
we obtain the final result for the statistical density matrix 
\begin{equation}
	\rho = \exp\{\Omega - \beta ( \hat{H} - \mu \hat{N} 
	- A \hat{I}^{0} ) \}. \label{final-DM}
\end{equation}
This is the general form, valid even for a fully interacting system. 
It is an interesting observation, that a sole requirement 
of the constraint on the time average of current operator 
is equivalent to the strong stationarity of the DM and the constraint on
the current operator. 

\begin{figure}
\onefigure[width=6cm]{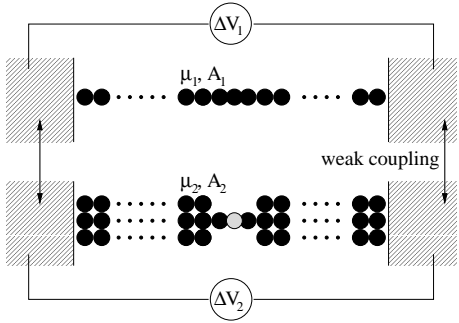}
\caption{Compound system 1+2 (see text) and related quantities.}
\label{Fig-1}
\end{figure}

We will now deduce the meaning of the parameter $A$. Consider we have two 
steady-state systems $1$ and $2$, with nonzero currents 
(see Fig.\ref{Fig-1}) described by their respective DM's $\hat{\rho}_{1(2)}$. 
The current is maintained with the parameters $A_{1(2)}$, 
but we can equally well imagine, that far left and far right there are ideal 
reservoirs to which we apply bias $\Delta V_{1(2)}$, so that $A$ and 
$\Delta V$ is in correspondence. If we were to describe a single compound 
system, comprising weakly coupled systems $1$ and $2$, with only the total
current being known, the DM would have had the form 
$\rho_{1+2} = \exp\left\{\beta \left( H_{1}+H_{2} - \mu(N_{1}+N_{2}) - 
			A (I^{0}_{1} + I^{0}_{2}) \right) \right\} $. 
On the other hand, if we weakly couple the originally disconnected $1$ and $2$,
we have $\hat{\rho}_{1,2} = \hat{\rho}_{1} \hat{\rho}_{2}$.
Clearly, the averages with respect to $\rho_{1+2}$ and $\rho_{1,2}$ will 
be the same if $\beta_{1}=\beta_{2},\mu_{1}=\mu_{2}$ and $A_{1}=A_{2}$, {\it
i.e.} no change of total current, as well as of $I_{1}$ and $I_{2}$, is 
introduced by coupling. Exactly this happens when the applied biases 
$\Delta V_{1(2)}$ 
are identical, so we conclude that $\Delta V$ should be a universal 
function of $A, \mu$ and $\beta$. The universality comes from the fact that 
systems $1$ and $2$ are arbitrary. The `$\Delta V$-meter' could be represented 
by a Landauer's concept of infinitely large reservoirs adiabatically connected 
through 1D conducting channel~\cite{Landauer90}. We leave detailed analysis 
of this situation for a future paper and here infer the $A-\Delta V$ 
relation from specific results in the following, giving 
$\Delta V = 2\tilde{A}$ for small $\tilde{A}$. This is a very general 
thermodynamical statement and removes the detailed considerations of 
near-equilibrium in reservoirs from the actual transport problem of interest 
in the nano-contact. 

We will now demonstrate several features of the general theory 
developed above, at the level of a self-consistent single-particle 
approximation. In the single-particle approximation,  is sufficient to know
the single-particle density matrix for evaluation of any quantity,
in our case the current and the electron density. These are well-defined 
for infinite system, unlike the total energy or total number of particles.
Since we deal with a system that is genuinely infinite, {\it i.e.} there
is a potential drop when comparing the right and left asymptotic regions,
with uniform current flowing, we need to resort to Matsubara Green's 
function techniques to obtain the density matrix unambiguously. The result is
\begin{equation}
	n(x,x') = \label{local-Fermi} \int dE \sum_{\alpha}
	\frac{\chi_{E,\alpha}(x)\chi_{E,\alpha}^{*}(x')}{
	\mathrm{e}^{\beta(E - \tilde{A}I_{\alpha}(E) - \mu)} + 1},
\end{equation}
a Fermi-like distribution with the effective dispersion 
$\tilde{E}(k)=E - \tilde{A}I_{\alpha}(E)$.
The $\chi_{E,\alpha}(x)$ diagonalise the effective Hamiltonian $\hat{K}$.
$I_{\alpha}(E)= \pm |t(E)| \sqrt{\frac{\kappa}{k}}$ are the eigenvalues
of the invariant current operator, which in the basis of right- and left-
going energy normalised scattering states has the form
\begin{eqnarray}
        \label{current-matrix}
        2 \pi \mathbf{I}^{0}(E) = \left[ \begin{array}{l l}
                t^{*}t \frac{\kappa}{k}
                & - r^{*} \tilde{t} \sqrt{\frac{k}{\kappa}} \\
                -r \tilde{t}^{*} \sqrt{\frac{k}{\kappa}}
                & - \tilde{t}^{*} \tilde{t} \frac{k}{\kappa}
                \end{array} \right].
\end{eqnarray}
The states $\chi_{E,\alpha}(x)$ are unitary transformation of the 
scattering states given by the eigenvectors of the matrix $\mathbf{I}^{0}(E)$
at each energy level $E$. $t,r$ and $\tilde{t},\tilde{r}$ are the usual
forward and backward transmission and reflection coefficients respectively,
and finally $k = \sqrt{2E}$ and $\kappa = \sqrt{2(E + \Delta \phi)}$ 
with $\Delta \phi$ being the drop in electrostatic potential energy.
Crucially, the scattering states appear here just as a convenient 
complete set of eigenstates of the Hamiltonian and it is the states
$\chi_{E,\alpha}$ which are actually being occupied according to 
Fermi-like occupancies in Eq.~(\ref{local-Fermi}). In the limiting case 
of $|r(E)| \rightarrow 0$
we obtain the original right- and left- going scattering states, in 
agreement with the occupation scheme. On the other hand, for
$|t(E)| \rightarrow 0$ we get nearly their symmetric and antisymmetric 
combinations. We discuss the physical significance of these in later 
paragraphs.

Next we give our motivation for the identification of $\tilde{A}$ as the 
applied bias through $\Delta V = 2\tilde{A}$. We look at the expectation value 
of the current operator in a 1D perfect wire. In the small $\tilde{A}$ limit 
we have
\begin{eqnarray}
	I = 2 \sum_{\alpha=0,1} \int_{0}^{\infty} 
                \frac{dE/2\pi}{\mathrm{e}^{\beta(E - \tilde{A}I_{\alpha}(E) 
                - \mu)} + 1} 
             (-1)^\alpha | t(E) | \sqrt{\frac{\kappa}{k}}  
        = \frac{2 e}{h} 2 \tilde{A} |t(E_{F})|^2. \label{landauer}
\end{eqnarray}
Since it is an experimentally well-established fact that the 
conductivity of a 1D channel is $2e^2/h$~\cite{vanWees},  we can directly 
identify $2\tilde{A}$ with the bias applied between two equilibrium reservoirs.
Due to the general arguments above we know that this relation is universal 
(for small $\tilde{A}$), so it needs to have the same form for any system. 
Eq.~(\ref{landauer}) is in complete agreement with Landauer's
formula~\cite{Buttiker85} even though it comes from rather different 
considerations.

In the following we will be concerned with the self-consistent determination 
of the drop in electrostatic potential $\Delta \phi$, and a detailed
discussion of the difference between $\Delta \phi$ and the applied bias
$\Delta V$. Specifically, let us suppose that our system consists of 
two identical $D$-dimensional jellium-like leads. Local neutrality 
requires ($\beta \rightarrow \infty$):
\begin{equation}
	\int_{0}^{\mu} \frac{dE}{2 \pi} \left[ n(x\rightarrow -\infty) 
	- n(x \rightarrow \infty) \right] 
	= \int_{-\Delta \phi}^{0} \frac{d \mathbf{k}^{D}}{ (2\pi)^D }.
\end{equation}
The meaning of this is that the charge appearing below the potential drop,
on the right, must be
exactly compensated by the charge transfered to the left by means 
of the occupancies in Eq.~(\ref{local-Fermi}). We can analytically 
evaluate the left-hand side for small $\tilde{A}$, obtaining  
\begin{equation}
\Delta \phi =  2 \tilde{A} |t|^2 |r|^2 = \Delta V |t|^2 |r|^2 \label{drop-A},
\end{equation}
independent of dimensionality $D$. Through this we can relate the 4-point 
conductance $G_{4P}=I/\Delta \phi$ to the 2-point conductance 
$G = I/\Delta V$. We immediately see, that the former gives a surprising 
result $G_{4P} = \frac{2e^2}{h} \frac{1}{|r|^2}$, approaching the quantum of
conductance for $|t| \rightarrow 0$. This contra-intuitive result
can be understood in terms of  
the occupation of nearly anti-symmetric admixtures of right- and left- going 
scattering states, present in $\chi_{E,+}(x)$. While this comes out of our 
formalism, we can expect that these combinations for the weakly connected 
system will be destroyed by a finite lifetime of the single-particle states,  
arising from whatever weak scattering by phonons or other electrons.
If we model this fact by cancelling the off-diagonal terms in the 
invariant current matrix Eq.~(\ref{current-matrix}), the resulting 2-point 
conductance turns out to be $\tilde{G} = \frac{2e^2}{h}|t|^4$ and the 4-point 
conductance $\tilde{G}_{4P} = \frac{2 e^2}{h}\frac{|t|^2}{|r|^2}$, while 
the relation in Eq.~(\ref{drop-A}) remains unchanged. $\tilde{G}_{4P}$ 
obtained without 
the off-diagonal terms is in complete agreement with the seminal work of 
B\"{u}ttiker {\it et al.}~~\cite{Buttiker85}, while the 2-point formula
gives the conductance smaller by factor $|t|^2$. We would also like to stress,
that the off-diagonal elements could possibly play role for situations, when 
$|t| \sim 1$ and therefore lead to higher conductances than those obtained 
from the Landauer 2-point formula. 

In order to elaborate the relation between the maximum entropy theory and
OS we notice, that even though we work with only one parameter related 
to the number of particles, $\mu$,  
from~(\ref{local-Fermi}) we see, that we can define two auxiliary 
Fermi energies $\mu_{\pm}$ up to which the states $\alpha = \pm$ 
are occupied from $\mu_{\pm} - \tilde{A} I_{\pm}(\mu_{\pm}) = \mu$ 
\begin{figure}
\onefigure[width=4cm]{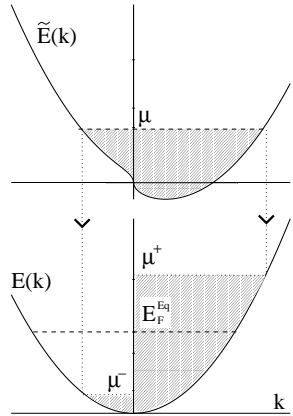}
\caption{The effective ($\tilde{E}(k)$) and the true energy ($E(k)$) dispersion
relations with the corresponding Fermi energies $\mu,\mu_{+}$ and $\mu_{-}$.}
\label{Fig-2}
\end{figure}
(see Fig.~\ref{Fig-2}). In the linear response we get 
$\Delta \mu = \mu_{+} - \mu_{-} = 2 \tilde{A} |t|$ which together 
with Eq.~\ref{drop-A} results in $\Delta \phi = \Delta \mu  |t| |r|^2 $.
Similarly, without the off-diagonal elements we have $\Delta \mu = 
2\tilde{A} |t|^2$ and  $\Delta \phi = \Delta \mu |r|^2 $. The latter 
relations demonstrate most clearly the difference between the maximum entropy
and OS. Firstly, when ignoring the off-diagonals, the applied 
bias $\Delta V$ in the OS is heuristically identified with $\Delta \mu$ while 
in our treatment the thermodynamical arguments given in the first part 
of this letter suggest $2\tilde{A} = \Delta \mu / |t|^2$.
Second, the right- and left- going states  are not unexceptionally
occupied by the left and right reservoir respectively; we believe that 
particularly for 
$|t| \sim 1$ can this effect be experimentally verified based on 
the differences between conductances coming from these two approaches. 

In conclusion, we have shown how the maximum-entropy formalism 
can be applied for non-equilibrium steady states. We have derived 
the statistical density matrix introducing the $\hat{L}$-operator
that guarantees the steady-state character of the statistical ensemble
and identified its resolution with Kubo's invariant part of the 
current operator. A Lagrange
multiplier $2\tilde{A}$, conjugate to the current operator, represents
the applied bias. In the second part of the paper we have demonstrated 
the theory on simple examples, discussing in detail the character of 
the density matrix within the single-particle approximation. We have shown 
that for systems with no reflection 
probability our theory gives results identical to the usual occupation
scheme. This agreement slowly breaks down as the transmission is decreased,
as the relevant states became a coherent combinations of right- and left- going 
states. We have derived a simple dimensionality-dependent formula
for the electrostatic potential drop and discussed its relation
to the applied bias within the context of our theory. 

\acknowledgments
The authors gratefully acknowledge useful discussions with
Carl-Olof Almbladh and Ulf von Barth. 
This work was supported by the RTN programme of the European Union 
NANOPHASE (contract HPRN-CT-2000-00167).

\end{document}